\documentclass[openacc]{rstransa}
\usepackage{comment}
\usepackage{bm}
\usepackage{bbm}
\usepackage{upgreek}
\usepackage{siunitx}

\newcommand{\cc}[1]{#1}

\titlehead{Research}

\begin{document}

\title{Practical primary thermo- metry via alkali-metal\cc{-vapour} Doppler broadening}

\author{N. Agnew$^{1}$, V. Vohn\'ikov\'a$^{1}$, E. Riis$^{1}$, G.~Machin$^{2}$, and A. S. Arnold$^{1}$}

\address{$^{1}$Department of Physics, SUPA, University of Strathclyde, 107 Rottenrow, Glasgow G4 0NG, UK\\
$^{2}$National Physical Laboratory, Hampton Rd, Teddington TW11 0LW, UK}

\subject{atomic physics, photonics, thermal dynamics}

\keywords{primary thermometry, photonic thermometry, Doppler broadening, atomic spectroscopy}

\corres{N. Agnew\\
\email{nicola.agnew@strath.ac.uk}\\ 
A.~S.~Arnold\\
\email{aidan.arnold@strath.ac.uk}}

\begin{abstract}
Doppler-broadening thermometry (DBT) can be used as a calibration-free primary reference suitable for practical applications, e.g.\ reliably measuring temperatures over long periods of time  in environments where sensor retrieval \cc{for recalibration} is impractical. We report on our proof-of-concept investigations into DBT with alkali metal vapour cells, with a particular focus on both absorption and frequency accuracy  during scans. We reach sub-kelvin temperature accuracy, and experimental absorption fit residuals below $0.05\,\%$, in a simple setup. The outlook for portable, practical devices is bright, with clear prospects for future improvement.
\end{abstract}

\begin{fmtext}
\section{Introduction}

In 2019, the General Conference on Weights and Measures culminated years of scientific progress by finalising the redefinition of the International System of Units in terms of fundamental constants  
\cite{Stock2019}. This means that the kelvin, which was previously anchored to the triple point of water,  is now defined in terms of a fundamental constant of nature -- the Boltzmann constant \cite{Machin2018}.

This change in the kelvin definition offers exciting new possibilities for realising and disseminating the kelvin directly, especially through practical primary thermometry. Traditionally the practical realisation of the kelvin has been indirect, i.e. through the defined scales, such as the International Temperature Scale of 1990 (ITS-90). Practical thermometers were calibrated traceably to the ITS-90 however these are prone to drift in use, thereby requiring periodic calibration, or sensor replacement to maintain traceability.  Correcting  for this sensor drift, for  
\end{fmtext} 
\maketitle

\noindent \cc{example} in industry, is essential to maintain process optimisation to ensure minimum energy use, emissions and zero-waste manufacturing.

The introduction of the \textit{mise-en-pratique} for the definition of the kelvin (\cc{\textit{MeP}}-K-19)  \cite{Fellmuth2016,Machin2022} has opened the path for the scientific and metrology communities to develop practical primary thermometers capable of directly realising the kelvin in process. Advances have been made in this area using primary thermometry techniques such as Johnson noise thermometry \cite{Bramley2016,Bramley2020}, acoustic gas thermometry \cite{dePodesta2010} and Doppler broadening thermometry (DBT) \cite{Agnew2024,Machin2025}. Recent research in these areas was initially developed so as to determine a low-uncertainty value for the Boltzmann \cc{constant}, in support of the kelvin redefinition. This required complex, high-precision experimental setups \cite{Feng2017,Podesta2013, Moretti2013}.
However, subsequent research has focused on reducing the complexity in an attempt to produce practical primary thermometers that combine the usability of secondary thermometers but with the long-term stability offered by primary thermometers. Such devices provide direct traceability to the kelvin at the point of measurement and as such require no calibration, which instead passes to the measurement system. This means that practical primary thermometers should, in principle, give correct truly thermodynamic temperature readings for the lifetime of the sensor which is an important milestone in thermometry.

Our research is centred on DBT \cite{Gianfrani2016,Gravina2024}. This method employs spectroscopy on atomic or molecular species to measure the frequency broadening of transitions between specific energy levels, a phenomenon resulting from the thermal motion of the interrogated atomic/molecular species. Doppler broadening is directly linked \cc{to thermodynamic} temperature via the Boltzmann constant. While seemingly straightforward and thus well-suited for application as a practical device, the current approaches to the technique are far from simple. 
Critical requirements for successful implementation of DBT are outlined in Ref.~\cite{Gianfrani2016}, including the necessity for a highly reproducible frequency axis, requiring highly linear detectors and detection electronics, and an accurate lineshape model for precise temperature extraction from the spectroscopic data. 

Our DBT study focuses on alkali atoms, like other pioneering work in the area \cite{Truong2011,Truong2015,Truong2015NC,Pan2019}, specifically rubidium. Rubidium (Rb) offers advantages, such as its wavelength of $780\,$nm, where laser technology and linear detectors are readily available.  In addition, like all alkali metal species, Rb benefits from having a relatively simple energy level structure and lineshape model compared to its molecular counterparts, and extrapolation to zero pressure is less critical \cite{Tennyson2014,Gianfrani2016}. However, alkali atoms do not come without drawbacks -- they are sensitive to both magnetic fields and optical pumping effects, with the degree of sensitivity varying between transitions as well as species. For instance, the non-alkali metal strontium has emerged as a promising candidate for DBT due to its reduced susceptibility to magnetic fields with low systematic Zeeman effects \cite{delAguila2018}. 

Our work addresses, arguably, one of the most critical challenges: the need for a precise and well-defined frequency axis, without which identifying the Doppler half-width reliably is impossible. The high non-linearity associated with laser frequency scanning complicates the accuracy of DBT measurements. While etalons serve as effective `frequency rulers,' with resolution proportional to their length (itself ill-defined and liable to drift), the size required  for sub-MHz precision Doppler thermometry is prohibitively large. An alternative approach to obtaining a precise frequency axis is to use an optical frequency comb \cite{Truong2015,Truong2015NC}. \cc{However, while accurate, such an approach is bulky} and prohibitively expensive for practical portable applications.

To address this, we have developed a simple, tuneable, and phase-locked laser system that is directly traceable to an atomic frequency reference, with a relative linewidth less than $1\,$Hz. Here we demonstrate the effectiveness for frequency-calibrated DBT and highlight its architecture, which is designed for scalable integration with potential for compact on-chip devices. Notably, our device currently demonstrates temperature readings with sub-kelvin accuracy at room temperature, showcasing its reliability and utility in practical applications.

\maketitle

\section{Absorption thermometry vs Doppler-broadening thermometry}\label{Modeling the atomic lineshape}

For our study, we will primarily focus on the $^{87}$Rb $780\,$nm transitions from the 5S$_{1/2}$ $F_\textrm{g}$=1 ground state to the 5P$_{3/2}$ $F_\textrm{e}$=0, 1, 2 excited states. These three D2 hyperfine transitions constitute the Doppler feature that we examine experimentally. We use the Beer-Lambert law in conjunction with the sum of Voigt spectral profiles, convoluting the homogeneous Lorentzian broadening and inhomogeneous Gaussian Doppler-broadening, for each absorption line. Relative transition strengths are appropriately weighted, and we allow for the rapid change in alkali atomic vapour density with temperature. Full details can be found in Refs.~\cite{Foot2005,Siddons2008,Zentile2015,Keaveney2018,Pizzey2022,85RbSteck,87RbSteck,ADM}, and note that the theoretical approach is applicable to all ground-state-connected transitions in alkali metal atoms, including the D1 and D2 lines.

\begin{figure}[!b]
\centering\includegraphics[width=1\textwidth]{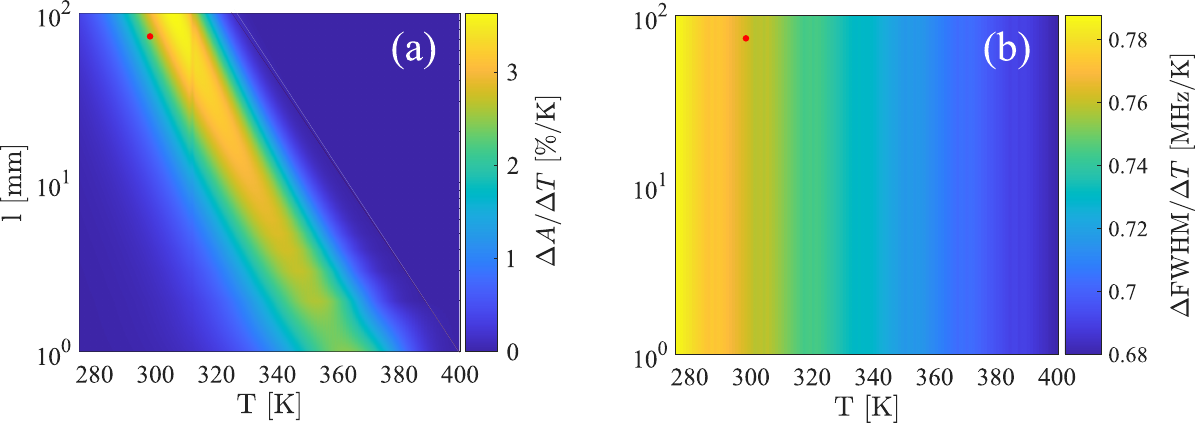}
\caption{Theoretical comparison of absorption and Doppler thermometry. (a) Variation in the peak absorption of the $^{87}$Rb 5S$_{1/2}$ $F_\textrm{g}$ = 1 to 5P$_{3/2}$ $F_\textrm{e}$ = 0, 1, 2 Doppler feature per kelvin temperature change, shown for different temperatures and cell lengths. (b) Change in the FWHM of the imaginary part of the electric susceptibility as a function of temperature and cell length. In this case, there is no cell length dependence. \cc{The red dots on both figures indicate the temperature and cell length chosen for the experimental realisation of DBT.}}
\label{87RbColourMaps}
\end{figure}

The model predicts changes in the Doppler features of alkali metal D-lines as a function of cell length and temperature. We first assessed how the maximum absorption of the considered transition varies per kelvin temperature change, as depicted in Figure~\ref{87RbColourMaps}~(a). \cc{We observe that the change in absorption with temperature is more pronounced at temperatures in the region of approximately $320\,$K - $370\,$K. However, as the optical density of the vapour increases, the transition begins to saturate, such that no light passes through the vapour at higher temperatures. The temperature at which saturation occurs also depends on the length of the vapour cell, with shorter cells reaching saturation at higher temperatures.} While it is feasible to estimate the temperature ad hoc by solely observing the absorption peaks of Doppler features \cite{Agnew2024}, this method is not considered primary due to the dependence of the absorption profile on the cell length. The data delineates a reasonable operational temperature range \cite{Truong2011} for a variety of cell lengths \cc{at natural isotopic abundance} \cite{Agnew2024}. \cc{For a cell length of $75\,$mm the operating range of this method is approximately $290\,$K to $340\,$K.}

Figure \ref{87RbColourMaps}~(b) demonstrates the change in the full width at half maximum (FWHM) of the atomic susceptibility profile as a function of temperature and cell length. Although the length axis is shown, it is included only to confirm the elimination of any length dependence. In this context, cell length merely acts as an amplitude scaling factor after the natural logarithm of the transmission is taken, underscoring the primary nature of this temperature sensing method. The change in FWHM is relatively uniform, approximately $0.76\,$MHz per kelvin, and indicates the precision required during a frequency scan over a Doppler feature to accurately derive  temperature at the kelvin level. Note, however, that temperature sensitivity of the absorption is a relatively good proxy for the signal-to-noise ratio of the DBT signal, so although the Doppler width is cell-length independent, Figure~\ref{87RbColourMaps}~(a) is a good companion to Figure~\ref{87RbColourMaps}~(b), to indicate regions of high DBT accuracy. For this reason we provide data near $297\,$K, with a $75\,$mm long $^{87}$Rb-enhanced vapour cell, indicated by the \cc{red} dot in both figures.

\section{Experimental procedure}

\begin{figure}[!b]
\centering\includegraphics[width=1\textwidth]{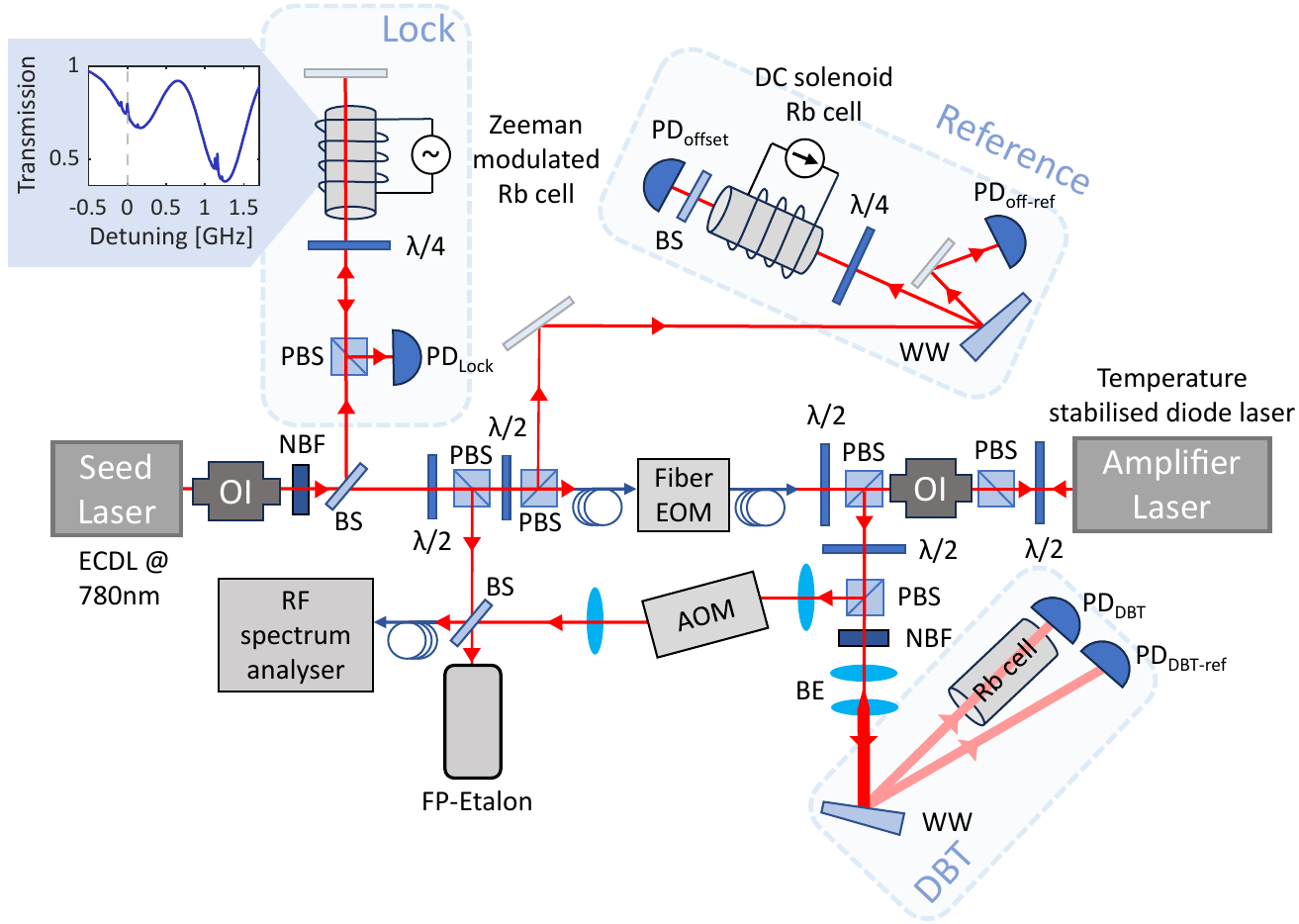}
\caption{Experimental setup. Abbreviations used: Half-wave plate ($\lambda/2$), photodiode (PD), acousto-optic or electro-optic modulator (AOM or EOM), Fabry-P\'erot etalon, (FP-Etalon) optical isolator (OI), [polarising] beamsplitter ([P]BS), narrow-band filter (NBF), wedged window (WW), external-cavity diode laser (ECDL). \cc{The PBSs shown next to the amplifier laser OI are the OI's internal PBSs.}}
\label{experimental_setup}
\end{figure}

The optical setup consists of a two-laser system, as shown in Figure~\ref{experimental_setup}. An external cavity diode laser (ECDL \cite{Arnold1998}) serves as the seed laser (SL), which is frequency-locked to the 5S$_{1/2}$ $F_\textrm{g}=2$ $\leftrightarrow$ 5P$_{3/2}$ $F_\textrm{e}=2,3$  crossover transition of an $^{87}$Rb hyperfine pumping spectroscopy signal \cite{Smith2004}  (Fig.~\ref{experimental_setup}, inset, dashed line).  
Frequency stabilisation is achieved using a lock with either laser-frequency or Zeeman modulation. The latter uses an oscillating magnetic field applied to the locking cell, which induces modulation of the hyperfine transition frequency rather than directly modulating the laser frequency, thereby minimising laser linewidth.

The locked SL is directed through a fibre electro-optic modulator\footnote{IXblue NIR-MPX800-LN-10 EOM driven by a SynthHD 10MHz – 15GHz RF Signal Generator} 
(EOM) \cc{-- tuneable from DC up to $16\,$GHz --} 
which generates optical sidebands at the modulation frequency ($f_\textrm{m}$). The output of the EOM is then injected into a temperature-stabilised  laser diode, referred to as the amplifier laser (AL). The centre frequency capture range of the AL is tuned by adjusting the laser diode's current. This is due to the bare  diode (with a typically imperfect anti-reflection coating) acting as an optical cavity, where changes in the injection current alter the effective cavity length, and consequently the centre frequency of the capture range. Additionally, the bandwidth of the capture range is found to be proportional to the ratio of the optical power of the SL injected into the AL 
to the output power of the AL. 
By carefully selecting the capture range, the carrier and any undesired EOM sidebands can be suppressed whilst amplifying the desired mode.  This method \cite{Agnew2024laser} provides a well-defined  optical frequency source referenced to an atomic transition with offset frequency determined by an RF generator, making it well-suited for DBT. It enables accurate frequency axis calibration -- ensuring precise spectroscopic measurements of the Doppler width.

\subsection{Characterising the seed laser lock}

Our frequency-offset AL is phase-locked to the SL, and thus has virtually no relative frequency linewidth. Consequentially, the frequency noise of both lasers is tied to SL frequency noise, which should therefore be characterised and minimised, as it is an important limitation to DBT measurement precision.  

Since our ultimate goal is to construct a portable, practical and accurate sensor, simplicity is key for all system elements -- including the SL lock. 
We have quantitatively compared the frequency noise of two relatively simple and inexpensive dither-locking techniques: ECDL cavity length frequency modulation via its built-in piezo-electric transducer, and Zeeman modulation of the lock vapour cell's atomic transition with a solenoid. To have a like-for-like comparison between the two locking techniques, we adjusted the error (derivative) signals to be of similar amplitude and slope for the peak we are locking to. We modulate both systems at a frequency $35\,$kHz with a sub-MHz peak-peak frequency modulation depth. When measuring the frequency noise of the Zeeman-modulation locked laser the built in piezo-electric transducer is only used for correcting drift in frequency, and when measuring the Piezo-modulation locked laser the coils around the vapour cell are disconnected from their power supply.

\begin{figure}[!t]
    \centering
    \includegraphics[width=\linewidth]{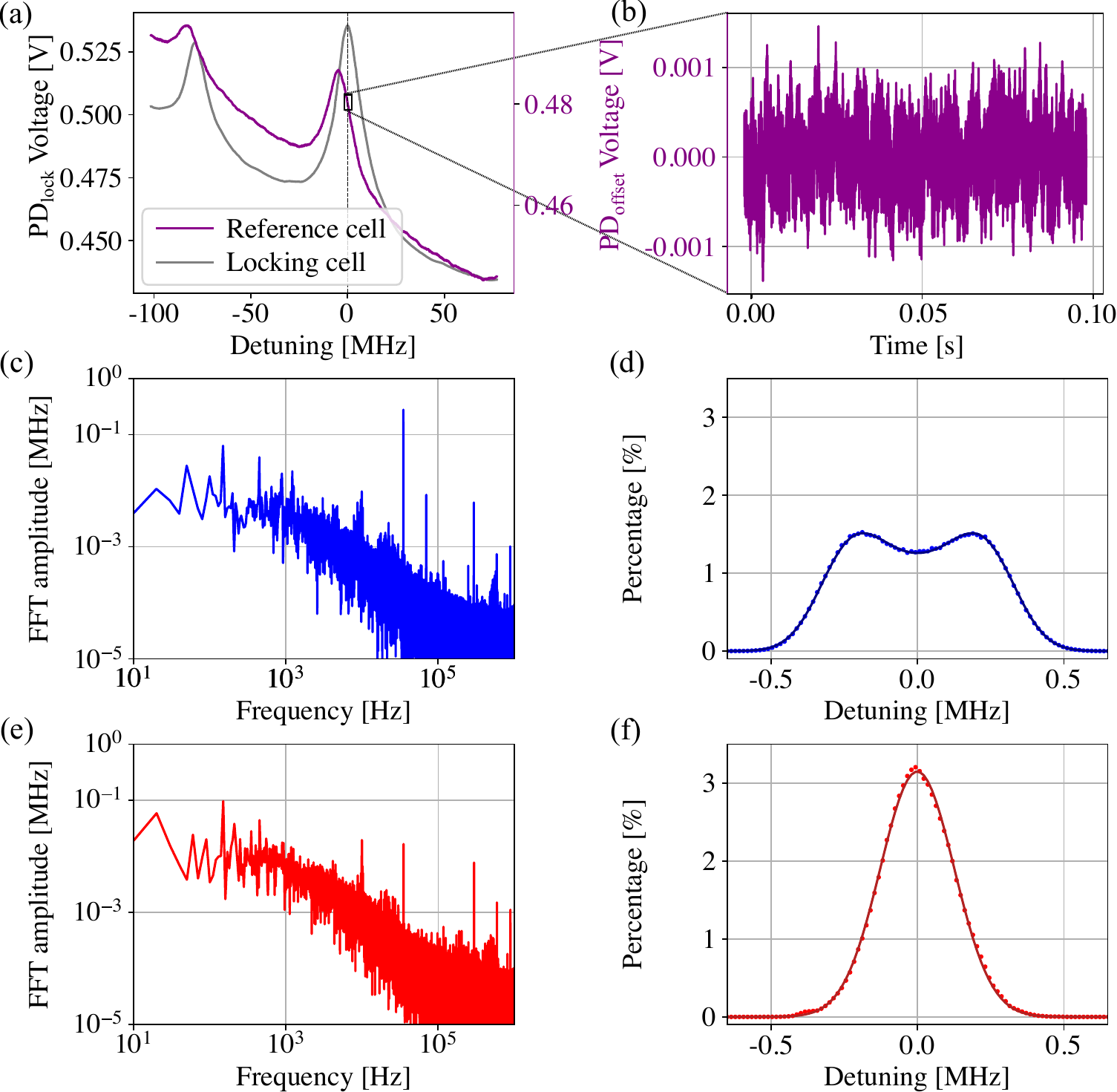}
    \caption{Locked-laser frequency stability: using the reference cell to compare between Piezo and Zeeman modulation (for setup see Figure~\ref{experimental_setup}). (a) Spectroscopy of the lock (PD$_\textrm{lock}$) and reference cell (PD$_\textrm{offset}$)  photodiode voltages, for frequency slope calibration. (b) Normalised reference cell PD trace (intensity-normalised with PD$_\textrm{off-ref}$) using a laser locked with Zeeman modulation ($0.1\,$s scan). FFTs of a $0.1\,$s frequency noise measurement of a Piezo (blue) and Zeeman (red) modulation locked laser \cc{are in (c) and (e)}. Frequency histograms with fits for a $1\,$s frequency noise measurement of Piezo (blue) and Zeeman (red) modulation locked laser \cc{are in (d) and (f)}. \cc{The fits are a Gaussian distribution convoluted with sinusoidal motion (Piezo) and a Gaussian distribution (Zeeman). See also the available datasets.}}
    \label{laser-stability}
\end{figure}

As the lock error signal (not shown) is linear in frequency near the lock point, it would provide a simple way to measure frequency noise -- however it is low-pass filtered, and may retain traces of both the modulation and any harmonics, so it is 
a poor discriminant of lock noise. To get an independent locked SL frequency noise signal with long-term accuracy, an additional vapour cell for separate hyperfine pumping spectroscopy was used \cite{willis1995,Daffurn2021,Singh2025}, analogous to the one for locking the laser, albeit single-pass and with a reference photodiode for intensity normalisation (dashed box labeled `Reference' in Figure~\ref{experimental_setup}). This  section of the setup used light sampled by a PBS prior to the fibre EOM, and the reference vapour cell uses a similar solenoid to the locking vapour cell but with a constant magnetic field. Due to the Zeeman effect, the cell's atomic transitions are shifted along the frequency axis (Figure~\ref{laser-stability}~(a)). The setup's field and quarter-wave plate were calibrated to yield the most linear and steep photodiode slope possible around the laser locking frequency, where the locking cell's photodiode slope is zero (dashed vertical line through $0\,$MHz in Figure~\ref{laser-stability}~(a)). 

First, a calibration trace was taken (Figure \ref{laser-stability}~(a)) and the linear slope of the stability reference cell peak calculated for a range $\approx 1\,$MHz centred on zero detuning of the locking frequency. Next, the laser was locked to the top of the crossover peak and the transmitted signal through the stability reference cell was recorded for three different time scales -- $0.1\,$s, $1\,$s and $100\,$s. 
Before any analysis was performed, the signal from the PD$_\textrm{offset}$ was divided by the normalised signal from the PD$_\textrm{off-ref}$ to filter out any intensity noise.
Fast Fourier transforms (FFTs) and frequency histograms of the data were made, with examples for a $0.1\,$s sample length FFT and a $1\,$s sample length histogram shown in Figures~\ref{laser-stability} (c) and (d) for Piezo-modulation locking; and (e) and (f) for \cc{equivalent} Zeeman-modulation locking, respectively. Root-mean-square (RMS) frequency noise could then be found from the FFT directly or via the standard deviation of the histogram fit, with very little difference between the two methods.

The motivation behind performing FFTs was to uncover and thereby minimise the main contributors to frequency noise (e.g.\ \cc{power line} noise at $50\,$Hz or the modulation frequency $35\,$kHz). It was implemented only for the $0.1\,$s sample lengths to get \cc{the} best resolution of frequency noise up to $1\,$MHz.
The frequency noise RMS calculated using the FFT of $0.1\,$s scans yielded $(220\pm1)\,$kHz for the Piezo modulation locked laser and $(135\pm2)\,$kHz for the Zeeman modulation locked laser, with standard error calculated over 10 scans. Frequency histograms were made of all three time scales and resulting RMS values from histogram fits are summarised in Table~\ref{tab:table1}.

In Figure~\ref{laser-stability} (c) the example of an FFT plot for the Piezo modulation lock technique shows that the dithering (at $35\,$kHz) is the biggest contributor to the frequency noise. For the $100\,$s measurement, the frequency noise of the Piezo modulation locked laser appears to be reduced but only due to limitations in the sampling rate. Each photodiode trace consists of $10^6$ points regardless of the measurement duration, which means $100\,$s scans have no frequencies higher than $5\,$kHz contributing to the overall noise, as expected from the Nyquist theorem \cite{Austerlitz2003}. This is also apparent from the histogram plots for the cavity length modulation -- for $0.1\,$s (not shown) and $1\,$s (Figure~\ref{laser-stability} (d)) the frequency has a classical distribution of a Gaussian convoluted with sinusoidal motion, while for over $100\,$s (not shown) it matches the Gaussian distribution, in a similar manner to the Zeeman modulation (which is Gaussian at all three time scales).

\begin{table}[!t]
    \caption{Table of frequency noise measurements for the seed laser. Root-mean-square (RMS) values obtained from histogram fits and standard errors were obtained (statistically) from ten scans \cc{of $10^6$ data points over} each time duration.}
    \label{tab:table1}
        \begin{tabular}{|d|c|c|}
        \hline
         \mbox{Duration [s]} & \mbox{Piezo mod.~RMS [kHz]} & \mbox{Zeeman mod.~RMS [kHz]}\\
        \hline
         0.1 & $219 \pm 1$ & $135 \pm 2$ \\
        1.0 & $215 \pm 2$ & $133 \pm 4$ \\
        100.0 & $141 \pm 2$ & $133 \pm 3$\\
        \hline
        \end{tabular}
\end{table}

In summary, local Zeeman modulation creates much less frequency noise in a locked laser than laser frequency modulation. The RMS level achieved here of $(133\pm3)\,$kHz for $100\,$s is significantly better than the $(880\pm390)\,$kHz seen in Ref.~\cite{Truong2015NC}, especially as Gaussian laser noise adds in quadrature to the Gaussian Doppler broadening. Moreover, the laser is currently only locked with (slow) piezo feedback, and additional (fast) diode current feedback may further reduce laser linewidth.

\subsection{Characterising the amplifier laser output}

\begin{figure}[!t]
\centering\includegraphics[width=1\textwidth]{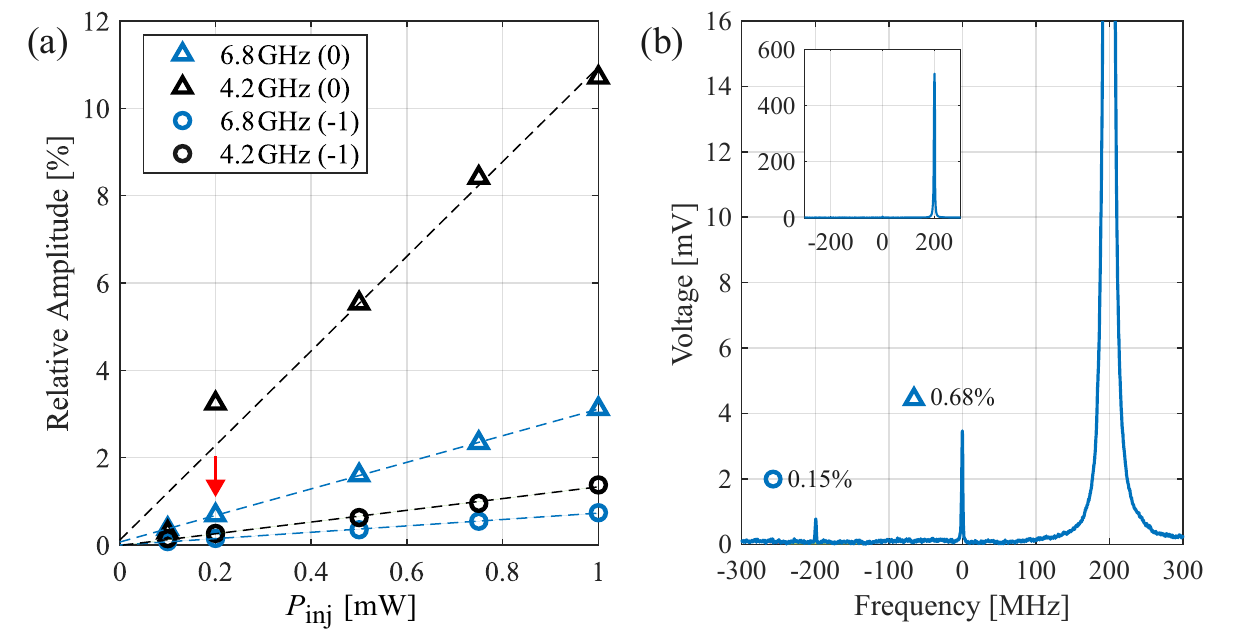}
\caption{(a) Amplitudes of the EOM's $0^\textrm{th}$ and -$1^\textrm{st}$ orders relative to the amplified $1^\textrm{st}$ order sideband across a range of injection powers at EOM modulation frequencies of $6.8\,$GHz and $4.2\,$GHz. (b) Example of a $1\,$GHz free-spectral-range Fabry-Pérot etalon trace at an EOM frequency of $6.8\,$GHz and an injection power of $0.2\,$mW, for the conditions  highlighted by the arrow in (a).}
\label{etalon_fig}
\end{figure}

We quantified the suppression of unwanted EOM orders by analysing the output of the AL using a Fabry-Pérot etalon. Figure~\ref{etalon_fig}~(a) shows the relative amplitudes of the -$1$ and $0$ rejected EOM orders relative to the amplified $+1$ order across a range of injection powers. The two selected EOM modulation frequencies for analysis correspond to representative frequency splittings between our SL lock point frequency  and the Doppler features of $^{85}$Rb 5S$_{1/2}$ $F_\textrm{g} = 2$ to 5P$_{3/2}$ $F_\textrm{e}$ = 1, 2, 3 at $\approx4.2\,$GHz, and $^{87}$Rb 5S$_{1/2}$ $F_\textrm{g} = 1$ to 5P$_{3/2}$ $F_\textrm{e}$ = 0, 1, 2 at $\approx6.8\,$GHz.  
An example of an etalon trace used to calculate the relative amplitudes is shown in Figure~\ref{etalon_fig}~(b). This trace was taken for an EOM frequency of $6.8\,$GHz for an injection power of $0.2\,$mW -- as indicated by the red arrow in Figure~\ref{etalon_fig}~(a). 

It is clear that we achieve a higher rejection ratio at lower injection powers, attributable to the decreased bandwidth of the AL's capture range. This also explains why we observe better rejection at higher EOM frequencies. Ideally, we want the capture range to be spectrally distinct from unwanted sideband orders to minimise unwanted amplification. This can be achieved by two methods: firstly, by increasing the modulation frequency of the EOM so that is is far from the carrier, hence our choice of the more widely separated $^{87}$Rb Doppler peaks, and we note Cs D-lines have $9.2\,$GHz separation making them even more suitable for DBT with this technique. 

\cc{A second} method for sideband rejection is by reducing the capture range, although this approach presents some challenges. If the bandwidth of the capture range is less than the width of the Doppler feature the AL laser current cannot be fixed as this prevents continuous scanning across the entire feature. This limitation is eliminated by implementing feed-forward control, allowing parallel scanning of the AL diode current and the EOM modulation frequency, however with a narrow capture range this necessitates the use of  a stable, low noise AL current driver \footnote{Thorlabs LDC201CU}. 
With our present current stability levels 
it is feasible to perform DBT scans around $6.8\,$GHz with an injection power of $0.2\,$mW and AL power of $32\,$mW.

The $0^\textrm{th}$ order poses a greater challenge than the -$1^\textrm{st}$ order, being spectrally closer to the amplified sideband. It should be noted that this data was acquired using an EOM modulation depth selected so that the carrier and $1^\textrm{st}$ order sidebands were of equal amplitude, to allow a fair comparison. However, although not used here, it is feasible to increase the modulation depth such that both first order sidebands are maximised while suppressing the carrier, \cc{use a fiber Bragg grating filter \cite{Macrae2021},} or use an additional fibre-EOM controlling amplitude to enhance single-sideband operation \cite{Dammalapati2025} prior to injection. 

As a second method of characterizing the amplifier's output, we used a radio frequency (RF) spectrum analyser. The amplified output was passed through an acousto-optic modulator (AOM), introducing a frequency offset of $80\,$MHz (Figure~\ref{experimental_setup}). The beat note between the locked SL and the $80\,$MHz offset AL was then measured. This offset allowed for a direct comparison of the relative amplitudes between the copied sideband and the suppressed sidebands. We observed an amplitude suppression of $\approx20\,$dB between the beat notes, indicating effective sideband rejection, and a beat note $-3\,$dB-width of $1\,$Hz \cite{Agnew2024laser}.

\subsection{Doppler broadening spectroscopy}

For DBT we scan over the $^{87}$Rb 5S$_{1/2}$ $F_\textrm{g}=1$ to 5P$_{3/2}$ $F_\textrm{e}=0, 1, 2$ Doppler feature. These overlapping Doppler transitions are selected because they are spectrally the furthest from our lock point, minimising noise from residual unwanted EOM sidebands, yet still easily remaining within the operating range of the fibre EOM. To reduce amplified spontaneous emission (ASE) noise from the lasers a $1.2\,$nm-wide narrow-band filter centred at $780\,$nm is placed immediately after the SL, which  is complemented by an additional $10\,$nm-wide narrow band $780\,$nm filter after the AL. This decreases laser ASE noise by up to a factor of 50, as demonstrated in Ref.~\cite{Agnew2024}. Excessive ASE can lead to systematic errors in the DBT signal by causing an apparent decrease in the absorption signal due to non-resonant light passing through the Rb cell. 

To enhance the signal-to-noise ratio we operate below even the lowest, stretched-state saturation intensity  of $^{87}$Rb (the $|F_\textrm{e}=2$, $m_F=\pm 2\rangle$ to $|F_\textrm{e}=3$, $m_F=\pm 3\rangle$ transition, with $I_\textrm{sat}=1.67\,$mW/cm$^2$). The probe beam is therefore expanded using a Galilean \cc{telescope, 
before} being sampled and split in two by an antireflection-coated wedged window (WW, Figure~\ref{experimental_setup}). The WW spatially separates the \cc{weak ($0.2\,\%$)} reflected beams to avoid any interference, and the AL beam has near-normal incidence on the WW to minimise any polarisation dependence in the splitting process. One WW beam passes through a $75\,$mm-long glass Rb vapour cell,  and the other WW beam is in free space. 

Both WW beams are then focused onto identical photodiodes (PD$_\textrm{DBT}$ and PD$_\textrm{DBT-ref}$), with the PD$_\textrm{DBT}$ voltage divided by the P$_\textrm{DBT-ref}$ voltage to enable spectroscopy with normalised transmission. This counteracts any laser power fluctuations caused by the AL current scan, creating a flat off-resonant transmission background. The last five edge points of the transmission scan dataset are then averaged, and used to define a transmission of $1$ (i.e.\ $100\,\%$),  to which the dataset is then  linearly scaled. 
The transmission data are then converted into absorption, proportional to the electric susceptibility, by taking the natural logarithm of the normalised transmission. This conversion removes the Doppler feature width dependence  on the physical cell length, reducing it to an amplitude scaling factor.
A complete scan consists of 100 data points, with photodiode voltages measured using a 6.5-digit multimeter\footnote{Keithley DMM6500}. 
An integration time of 5 power line cycles (NPLC = 5), equivalent to $100\,$ms, is used for each measurement. To ensure stable readings, a $1\,$s delay is introduced between each step, allowing sufficient time for the modulation frequency and diode current to stabilize before the measurement is recorded.

\cc{The Rb cell is enclosed in a mu-metal foil to minimize the influence of external magnetic fields. The residual magnetic field inside the shielding was measured using a Hall probe and found to be $<10\,\upmu$T in all directions. The temperature of the cell is monitored using two temperature sensors, the higher preforming of which has a repeatability of $150\,$mK and accuracy of $\pm100\,$mK in the range of $293\,$K to $323\,$K\footnote{Sensirion SHT85}}.

The DBT spectral fit consists of a sum of three Voigt profiles, 
computed using the Faddeeva function \cite{Schreier1992}, with the Lorentzian contribution fixed to the natural linewidth of the $^{87}$Rb D2 line at $6.065\,$MHz. The relative amplitudes of the transitions are determined by the square of the Clebsch-Gordan coefficients \cite{Louck2008} for each hyperfine level,  
while the frequency spacing between transitions is calculated using known hyperfine structure constants. 
A non-linear least-squares fitting algorithm is used to extract key physical parameters from the data. The fit includes three free parameters: temperature, absolute transition amplitude, and global frequency shift.

\section{DBT results}\label{Results and discussion}

\begin{figure}[!t]
\centering
\includegraphics[width=.7\textwidth]{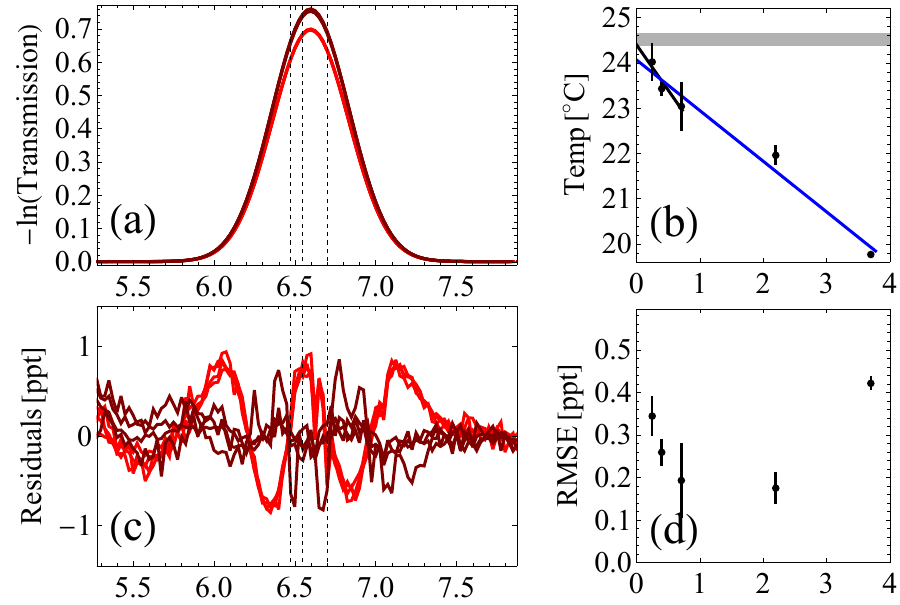}
\includegraphics[width=.7\textwidth]{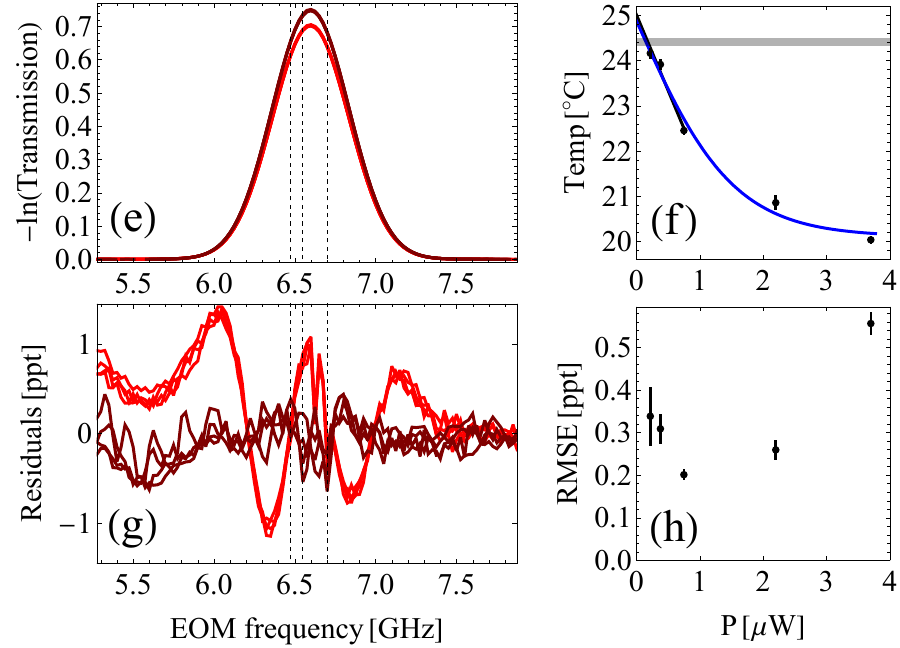}
\caption{\cc{Two different measurement runs (a)-(d) and (e)-(h), were taken with four measurements each at five different powers. The runs had opposite frequency scan directions (up and down, respectively), and different power orderings.  DBT scans for two AL powers, $0.7\,\upmu$W (dark red) and $3.7\,\upmu$W (light red) are shown in (a) and (e) -- with four overlapping sets of data and fits  for each power. Dashed vertical lines are the centres of the three overlapping Voigt profiles. Extracted temperature from the spectral fits as a function of AL intensity are shown in (b) and (f), with errors indicating standard deviations, and linear low-power (black) and tanh all-power (blue)  fits to the data. The corresponding fit residuals from (a) and (e) are shown in (c) and (g), respectively, in units of parts-per-thousand (ppt). The root-mean-square error (RMSE) of the fits as a function of AL power are in (d) and (h). A power of $1\,\upmu$W corresponds to a peak beam intensity $0.10\,$W/m$^2$, i.e.\ $0.60\%$ of the saturation intensity.}}
\label{DBT_fig}
\end{figure}

We performed DBT scans at various AL intensities, with a laboratory temperature of \cc{$\approx 24.5\,^{\circ}$C} as measured by the thermistors. Our goal was to investigate how laser intensity impacts the repeatability of the temperature measurements derived from the Doppler width. Representative Doppler scan \cc{data and fits from two  different data runs are shown in Figure~\ref{DBT_fig}~(a) and (e), respectively.} 
At low laser intensities, we found good agreement \cc{(Figure~\ref{DBT_fig}~(b) and (f))} between the DBT-extracted temperature and that measured by the thermistor installed in the cell environment. Specifically, we recorded \cc{extrapolated low-intensity} DBT temperatures within $0.5\,^{\circ}$C of the thermistor reading -- corresponding to a $\approx0.4\,$MHz frequency deviation. This indicates that, under low-intensity conditions, a relatively simple DBT setup may achieve sub-degree temperature differences when compared to standard secondary thermometry devices. Note that the temperature performance of the thermistor needs to be confirmed by calibration  -- but even uncalibrated, at room temperatures, they generally yield differences from the true temperature at the sub-kelvin level.

As the laser intensity increases, we observe both a reduction in absorption (Figure~\ref{DBT_fig}~(a), (e)) and a narrowing of the Doppler absorption feature (Figure~\ref{DBT_fig}~(b), (f)). This narrowing causes a systematic underestimation of the extracted temperature, akin to the behaviour observed in atomic power broadening -- but with the opposite \cc{effect.  
In contrast to the linewidth narrowing, we detected very little shift ($<0.1\,$MHz) in the centre frequency of the Doppler feature as a function of laser intensity -- as long as the cell's isotopic ratio of $98.5:1.5$ $^{87}$Rb:$^{85}$Rb was correctly accounted for}. 


\cc{Typical fit residuals at two powers are shown in Figure~\ref{DBT_fig}~(c) and (g). To fully} quantify the quality of the fits, we computed the root-mean-square error (RMSE) of the residuals across different laser intensities, as shown in Figure~\ref{DBT_fig}~\cc{(d) and (h)}. The RMSE reaches a minimum value of \cc{$0.2\,$ppt (parts-per-thousand) at powers of $\approx 1\,\upmu$W} \cc{- similar residuals have been reported using Cs \cite{Truong2015} when fitting with a Voigt profile only}. Above this threshold, the error increases nearly linearly with intensity. We attribute this trend to optical pumping effects, which become more prominent as the probing light intensity increases. This behaviour is clearly seen in the higher power residuals in Figure~\ref{DBT_fig}~(c) and (g).

Interestingly, the RMSE also increases at intensities below \cc{$\approx 1\,\upmu$W, likely due to a degradation} in signal-to-noise ratio at such low power levels. A potential solution for mitigating this issue in future implementations is to further expand the laser beam. This approach would allow operation at lower intensities while maintaining sufficient optical power, thus improving signal-to-noise performance without introducing significant optical pumping. The data for all powers were taken with identical termination at the photodiodes of \cc{\SI{1.2}{\mega\ohm}, and we intend to investigate how much the Figure~\ref{DBT_fig} results are affected by photodiode non-linearity, especially close to saturation}.

\section{Discussion and outlook}\label{Other broadening mechanisms}

\begin{table}[!b]
    \centering
\cc{    \begin{tabular}{ccccccccc}
Year & Species & $I$ & $\lambda$[nm] & $\eta_\Gamma$ & $N_\textrm{HF}$ & RSFR & Freq.~comb & Ref\\

\hline
2015 & $^{133}$Cs & $\frac{7}{2}$ & 895 & 79 & 1 & 100 & Yes & \cite{Truong2015,Truong2015NC} \\
2022 & $^{85,87}$Rb & $\frac{3}{2},\;\frac{5}{2}$ & 780 & 85, 84 & 3 & 4,000 & No & \cite{Pizzey2022} \\
2024 & $^{200}$Hg & 0 & 253 & 812 & 1 & 2,000 & Yes & \cite{Gravina2024} \\
2025 & $^{87}$Rb & $\frac{3}{2}$ & 780 & 84 & 3 & 200 & No & This work \\
\hline
\end{tabular}
\caption{
    A concise date-ordered comparison of metal vapour spectroscopy techniques capable of measuring Doppler-broadening. The ground level hyperfine system has 1 ($4I+2$) levels for $I=0$ ($I\neq0$ alkali-metal) nuclear spin, respectively, on a transition with wavelength $\lambda$, and a ratio of the Doppler:Natural widths at full-width-half-maximum of $\eta_\Gamma$ at $24.5^\circ$C. The number of unresolved hyperfine transitions (ignoring magnetic hyperfine) in each Doppler peak is $N_\textrm{HF}$. The estimated root-mean-square raw spectroscopy fit residue (RSFR) is given in parts-per-million -- and ours is $100$ if given relative to $100\%$ transmission rather than the $50\%$ absorption depth.}
    \label{compare}}
\end{table}

\cc{In its current iteration, the residuals of our direct Voigt DBT fitting routine compare favourably to the approaches used by other groups \cite{Truong2015,Truong2015NC,Pizzey2022,Gravina2024}. Moreover, we use a fairly simple approach without a frequency comb, and we address a relatively complex hyperfine transition, with a relatively low Doppler:Natural linewidth ratio (Table~\ref{compare}). In particular, we currently employ a Voigt-only fit, and are not reducing the residual by including either optical pumping or multiple etalons in the spectral modelling \cite{Truong2015NC}. The key residual seen in Fig.~\ref{DBT_fig} has the signature of optical pumping, which we hope to reduce by further expansion of the beam. 

Our simplified spectral model also}  does not fully account for all underlying broadening mechanisms, 
thus impacting the apparent temperature derived from the data. Presently, the model includes the two most significant effects: Gaussian Doppler broadening ($\approx500\,$MHz) and Lorentzian natural broadening ($\approx6\,$MHz). Any additional broadening mechanisms that increase the FWHM are attributed by the fitting routine to temperature-dependent Doppler broadening, resulting in an overestimated temperature reading, \cc{where $0.5\,^{\circ}$C is equivalent to $400\,$kHz.} 

\cc{A factor affecting the quality of the DBT fit is \textit{Zeeman broadening} which depends on both the magnitude of the magnetic field and its orientation relative to the laser polarisation. At low fields $<500\,\upmu$T (including typical geomeagnetic fields of $<70\,\upmu$T) the magnetic sublevels of alkali metal atoms, such as Rb, 
experience linear splitting proportional to their Land\'{e} g-factor and magnetic quantum number $m_F$. 
Given that our shielded total field is $<10\,\upmu$T we estimate any broadening to be less than $300\,$kHz. 
We will quantify the effect of Zeeman broadening on our experiment by introducing a controlled spatially uniform magnetic field so as to 
compare actual results to theory, and aid analysis of the associated systematic error. 
Such insights will be important to give an indication as to what extent future sensors must be shielded to avoid potential perturbation by strong magnetic fields.} 

Additional broadening mechanisms include \textit{Lorentzian self broadening}, arising as a result of binary collisions \cite{Lewis1980}. This effect becomes more pronounced at high temperatures due to increased number density. For rubidium at room temperature, the assumed number density is sufficiently low that collisional effects are marginal, typically less than $1\,$kHz. However\cc{, given the rapid permeation of atmospheric helium into regular glass cells, \textit{collisional broadening} of the Rb transition could be as high as $80\,$kHz \cite{Rotondaro1997}.} Both the Rb and background gas spectral broadening and shifts can be double-checked via DBT in \cc{He-protected }high-purity cold atom Rb vacuum cells \cite{Burrow2021}, or those with tuneable buffer-gas pressure \cite{Dyer2023}.  

\textit{Transit-time broadening} results from the atoms' finite interaction time with the passing beam \cite{87RbSteck}. A longer interaction time effectively reduces transit time broadening. For an interaction time corresponding to a $4\,$mm diameter 
-- the narrowest part of the beam axis -- the transit time broadening is $\approx50\,$kHz. This estimate is a highly simplified approximation that assumes atoms travel the shortest path across the beam, representing a worst-case scenario, \cc{and the effect could also be reduced with a larger beam}.

The intensity of the interacting light relative to the saturation intensity of Rb gives rise to \textit{power broadening}. As the light approaches the saturation intensity of the transition, the absorption near resonance significantly decreases, while the absorption far from resonance remains relatively unchanged \cite{Foot2005}, resulting in a broader FWHM. \cc{At $0.7\,\upmu$W} this effect is fairly marginal, contributing \cc{$\Gamma(\sqrt{1+I/I_\textrm{sat}}-1)\approx13\,$}kHz of Lorentzian-type broadening.  

\section{Conclusions}

We have presented our work focused on developing a practical and scalable primary thermometer employing DBT in alkali-metal vapour cells. Our Doppler-width measurements are facilitated by our frequency tunable, phase-locked amplifier laser system, with sub-Hz relative linewidth compared to the seed laser. This provides traceable frequency readings relative to an absolutely accurate atomic reference. \cc{We also} demonstrate Doppler extracted temperatures at room temperature consistent with thermistor readings at the sub-kelvin level in the limit of low probe-laser intensity. 
Future work will focus on developing a deeper understanding of the systematic effects that influence the reliability of DBT measurements \cc{-- intensity dependence, detector non-linearity, probe laser spectral purity, collisional shifts and Zeeman shifts -- as well as introducing improved and traceable thermometry}. Further, we aim to attempt DBT in smaller vapour cells \cite{Dyer202chip} to support miniaturisation efforts, and to extend our investigations across a broader temperature range to assess temperature measurement performance under varied operating conditions. 

\section{Acknowledgements}
We are grateful for insightful  discussions with Livio Gianfrani, careful reviewing by Sonja Franke-Arnold, Aldo Mendieta \cc{and our three anonymous reviewers}, and support via Engineering and Physical Sciences Research Council EP/T001046/1, EP/X525017/1; and National Physical Laboratory/EPSRC studentships 2749424 and 2931764. For the purpose of open access, the
authors have applied a Creative Commons Attribution (CC BY)
licence to any Author Accepted Manuscript (AAM) version
arising from this submission.

\bibliographystyle{RS}

\providecommand{\noopsort}[1]{}\providecommand{\singleletter}[1]{#1}%

\end{document}